\documentclass[twocolumn,superscriptaddress,aps,amsmath,amssymb,floatfix,showpacs]{revtex4}
\usepackage{graphicx}
\usepackage{bm}
\begin{document}
\title{Electron-phonon coupling and electron self-energy in electron-doped graphene:
calculation of angular resolved photoemission spectra.}
\author{Matteo Calandra}
\author{Francesco Mauri}
\affiliation{CNRS and Institut de Min\'eralogie et de Physique des Milieux condens\'es, 
case 115, 4 place Jussieu, 75252, Paris cedex 05, France}
\date{\today}

\begin{abstract}
We obtain analytical expressions for the electron self-energy
and the electron-phonon coupling in electron-doped graphene using
electron-phonon matrix elements extracted from density functional theory simulations.
From the electron self-energies we calculate angle resolved photoemission spectra.
We demonstrate that the measured 
kink at $\approx -0.2$ eV from the Fermi level is actually
composed of two features, one at $\approx -0.195$ eV due to the twofold degenerate
E$_{2g}$ mode, and a second one at $\approx -0.16$ eV due to the A$_{1}^{'}$ mode.
The electron-phonon 
coupling extracted from the kink observed in ARPES experiments is roughly 
a factor of 5.5 larger than 
the calculated one. This disagreement can only be 
partially reconciled by the 
inclusion of resolution effects. Indeed we show that a finite
resolution increases the apparent electron-phonon coupling by
underestimating the renormalization of the electron velocity 
at energies larger than the kinks positions.
The discrepancy between theory and experiments is thus 
reduced to a factor of $\approx$ 2.2.
From the linewidth of the calculated ARPES spectra we obtain the electron
relaxation time. A comparison with available experimental data 
in graphene shows 
that the electron relaxation time detected in 
ARPES is almost two orders of magnitudes smaller than what  
measured by other experimental techniques.

\end{abstract}
%%      71.15.Mb Density functional theory, local density approximation, 
%%               gradient and other corrections
%%      71.20.Tx Fullerenes and related materials; intercalation compounds
%%               Superconductivity:
%%      74.25.Jb Electronic structure
%%      74.25.Kc Phonons
%%      74.70.-b Superconducting materials  (for cuprates see 74.72.-h)
%%      74.70.Ad Metals; alloys and binary compounds (including A15, MgB2, etc.)
\maketitle

Despite the fact that the band structure of graphene 
is calculated in many solid-state textbooks \cite{Harrison,Wallace},
its experimental verification  
has been provided only recently by Angular Resolver Photo-Emission
(ARPES) measurements on a graphene monolayer deposed on a SiC substrate
\cite{Novoselov2005, ZhouPRB2005, Zhou2006NatPhys, Bostwick2007}.
The experiments show that the peculiar features of the electronic structure 
predicted theoretically are qualitatively confirmed by experiments: 
the carbon $\pi-$bands (i) cross at the K-point
in the Brillouin-zone (Dirac point) and (ii) depart linearly with a slope
$v_f$ from the Dirac point, (iii) the Fermi velocity extracted from experiments
\cite{Zhou2006NatPhys,Novoselov2005} is slightly larger (10-20\%) than that
calculated theoretically using density functional theory (DFT).

Beside this encouraging agreement between theory and experience, 
recent angular photoemission (ARPES) experiments  
\cite{Bostwick2007,Zhou_Private} 
performed on graphene revealed remarkable surprises.
Two kinks are seen in the ARPES dispersion: the first one
is at energies of 0.2 eV below the Fermi level ($\epsilon_f$) and its
energy position respect to $\epsilon_f$ is unchanged as a function of the 
doping level while the second one is closer to the Dirac point and 
its energy position respect to $\epsilon_f$ decreases rapidly
as the doping level is increased (see Fig. 2 in Ref. \cite{Bostwick2007}).
The first kink has been attributed to a phonon feature \cite{Bostwick2007}, 
while the second
kink has been interpreted as due to a plasmon\cite{Hwang2006,Bostwick2007}. 
In what follows we focus on the first kink. 

The ARPES momentum distribution curves (MDCs) associated to the -0.2 eV kink display
a puzzling behavior as a function of doping. Indeed it is observed that the 
magnitude of the jump associated to the MDC-linewidth in the 
$-0.5\, {\rm eV}< \epsilon-\epsilon_f < 0$ eV 
energy window decreases as a function of doping 
(see Fig. 3 in Ref. \cite{Bostwick2007}, where from top to bottom the 
jump increases). This is surprising since if this jump is associated to
the electron-phonon interaction then it should reflect the imaginary part
of the electron self-energy due to the electron-phonon interaction.
Since the magnitude of such interaction is usually
proportional to the density of states at the Fermi level, the jump
should increase as the doping level is increased.
Thus the opposite behavior should be expected.
This contradiction can be solved by noting that 
at low doping the tail of the second peak (attributed to a plasmon), 
is fairly close in energy and could affect the low energy part of the momentum
distribution curve. At larger dopings \cite{Bostwick_Review} the plasmon
peak has no effect, the electron phonon coupling does increase as a function
of doping. Thus we focus in the doping region identified
by $\epsilon_f > 0.3$ eV (the energy-zero being at the Dirac point). 

In this work we calculate the electron-phonon coupling parameter and the
electron-phonon coupling contribution to the electron self-energy in 
doped graphene. In particular, we give an explicit demonstration of
eq. 1 in ref. \cite{Calandra2005}. 
From the electron self-energy we obtain the spectral-weight function and
the ARPES spectra. Finally we compare the calculated spectra with available 
experimental data, discussing in the details the important finite-resolution
effects. 

The paper is structured as follows. 
In section \ref{sec:elph})
we obtain an analytical expression for the electron-phonon coupling in doped
graphene. The electron self-energy is calculated in sec. \ref{sec:selfenergy}.
The ARPES spectra are calculated from the electron self-energy in
sec. \ref{sec:results}, including finite
resolution effects. Finally in sec. 
\ref{sec:experimentals} we compare the electron
relaxation time measured by different experimental techniques
for both electron-doped graphene and graphite. Sec. \ref{sec:conclusions}
is devoted to conclusions.

\section{\label{sec:elph}Electron-phonon coupling in doped graphene}

In units of $2\pi/a$ with $a=2.46 {\rm \AA}$, the 2D volume of the graphene  
Brillouin-zone (BZ) is $\Omega=\frac{2}{\sqrt{3}}$.
The graphene $\pi^{*}$ bands are linear with slope $\beta=\hbar v_f=5.52 {\rm eV \AA}$
(within DFT) close to 
the Dirac points ${\bf K}=(1/3,1/3,0)$ 
 and ${\bf K}^{'}=2{\bf K}$. 
%The density of states per spin
%at the Fermi level $\epsilon_f$ 
%is $N_{\sigma}(\epsilon_f)=\frac{2\pi\sqrt{3}k_f}{\beta}$
%where $k_f=\epsilon_f/\beta$ is the Fermi momentum.
The density of states per spin at a general energy $\epsilon$ above or below
the Dirac point, but still in the region where the $\pi^{*}$ bands can
be considered linear, can thus be written as

\begin{equation}
N_{\sigma}(\epsilon)=\frac{2\pi\sqrt{3}|\epsilon|}{\beta^2}=
\frac{4\pi|\epsilon|}{\Omega \beta^2}.
\label{eq:dos}
\end{equation}
Note that in this work energies are always measured respect to the Dirac
point.

The electron-phonon coupling for a mode $\nu$ at momentum ${\bf q}$ 
due to the $\pi^{*}$ bands in graphene is
given by:
\begin{eqnarray}
\lambda_{{\bf q}\nu}&=&\frac{2}{\hbar\omega_{{\bf q}\nu}N_{\sigma}(\epsilon_f)}
\int_{BZ} \frac{d {\bf k}}{\Omega}
|g_{{\bf k} \pi^{*}{\bf k}+{\bf q} \pi^{*}}^{\nu}|^2\times \nonumber \\
&\times& \delta(\epsilon_{{\bf k}}-\epsilon_f)
\delta(\epsilon_{{\bf k}+{\bf q}}-\epsilon_f)
\label{eq:lambda_start}
\end{eqnarray}
where $\omega_{{\bf q}\nu}$ is the phonon frequency of the mode $\nu$ at
momentum ${\bf q}$ and  $g_{{\bf k} \pi^{*}{\bf k}+{\bf q} \pi^{*}}^{\nu}$ is the
electron-phonon matrix element for the $\pi^{*}$-band $\epsilon_{\bf k}$ and
for the phonon mode $\nu$.

To illustrate how the integral is evaluated we introduce the following
two regions of space, namely the sets:
\begin{eqnarray}
{\cal F}_{\bf K}(\epsilon)&=&
\{{\bf k }\,|\,\beta |{\bf k}-{\bf K}|< \epsilon +\eta\}  \\
{\cal F}_{\bf K^{'}}(\epsilon)
&=&\{{\bf k}\,|\,\beta |{\bf k}-{\bf K^{'}}|< \epsilon +\eta\}
\end{eqnarray}
In these definitions, $\eta $ is a small positive quantity.
For $\epsilon=\epsilon_f$,
since we assume that the Fermi level is not too far from the Dirac point
so that the $\pi^{*}$ bands are linear,
$|{\bf k}-{\bf K}|$ or $|{\bf k}-{\bf K}^{'}|$ is a small but finite vector
and ${\cal F}_{\bf K}(\epsilon_f)\bigcap {\cal F}_{\bf K^{'}}(\epsilon_f)$
is empty. 
The boundary of each region
of space at $\eta=0$ (circumference) is indicated as
$\partial{\cal F}_{\bf K}(\epsilon_f)$ and $\partial{\cal F}_{\bf K^{'}}(\epsilon_f)$.

In Eq. \ref{eq:lambda_start},
the two $\delta-$functions restrict the ${\bf k}$ integrations
to the region of space satisfying
the conditions $\epsilon_{{\bf k}}=\epsilon_f$ and 
 $\epsilon_{{\bf k}+{\bf q}}=\epsilon_f$.
The set of $\bf k$ points such that $\epsilon_{{\bf k}}=\epsilon_f$
is composed by the set
$\partial{\cal F}_{\bf K}(\epsilon_f) \bigcup \partial{\cal F}_{\bf K^{'}}(\epsilon_f)$.
%\includegraphics[width=0.9\columnwidth]{graphene_BZ_fs_smallef.eps}
%\caption{Simplified plot of the Graphene Brillouin zone and Fermi surfaces.}
%\label{fig:FS-graphene}
%\end{figure}
Thus in the integral in Eq. \ref{eq:lambda_start},
two cases are given (labeling ${\bf k^{'}}={\bf k}+{\bf q}$): 
\begin{itemize}
\item[(i)]{ ${\bf k}, {\bf k^{'}} \in 
\partial{\cal F}_{\bf K}(\epsilon_f)$ or ${\bf k}, {\bf k^{'}} \in
 \partial{\cal F}_{\bf K^{'}}(\epsilon_f)$,}
\item[(ii)]{  ${\bf k} \in \partial{\cal F}_{\bf K}(\epsilon_f)$ ,
 ${\bf k^{'}}\in \partial{\cal F}_{\bf K^{'}}(\epsilon_f)$
or vice versa}
\end{itemize}
In case (i) scattering occurs at ${\bf q}= {\bf \Gamma}+{\bf {\tilde q}}$, with 
small ${\bf {\tilde q}}$,
and $\pi^{*}$ bands can only couple to the twofold degenerate E$_{2g}$ phonon mode.
In case (ii) scattering occurs at ${\bf q}={\bf K}+{\bf {\tilde k}}$
or at ${\bf q}={\bf K}^{'}+{\bf {\tilde k}}$
, with small ${\bf {\tilde k}}$ , and 
$\pi^{*}$ bands can only couple to the A$_{1}^{'}$ phonon mode \cite{Piscanec2004}.
The electron-phonon matrix elements involved in the two scattering process
have been fitted to {\it ab initio} data in Ref. \cite{Piscanec2004} and are:
\begin{eqnarray}
|g_{{\bf K}+{\bf {\tilde k}}\pi^{*},{\bf K}+{\bf {\tilde k}}+{\bf {\tilde q}}\pi^{*}}^{E_{2g}}|^2 &=&
\langle g_{\bf \Gamma}^{2} \rangle 
[1\pm \cos(\theta_{{\bf {\tilde k}},{\bf {\tilde q}}}+
\theta_{{\bf {\tilde k}},{\bf {\tilde k}}+{\bf {\tilde q}}})] \label {eq:g_E2g}\\
|g_{{\bf K}+{\bf {\tilde k}}\pi^{*},{\bf K'}+{\bf {\tilde k}}+{\bf {\tilde q}}\pi^{*}}^{A_{1'}}|^2 &=&
\langle g_{\bf K}^{2} \rangle 
\left[1+\cos(\theta_{{\bf {\tilde k}},{\bf {\tilde k}}+{\bf {\tilde q}}})\right] \label{eq:g_A1}
\end{eqnarray}
In Eq. \ref{eq:g_E2g} the $\pm$ sign refers to the LO/TO E$_{2g}$ modes
respectively, $\langle g_{\bf \Gamma}^{2} \rangle = 0.0405 $eV$^2$ and
$\langle g_{\bf K}^{2} \rangle=0.0994$ eV$^{2}$, and $\theta_{{\bf u},{\bf v}}$ is
the minimal angle between the two vectors {\bf u},{\bf v}. 

For case (i) one has 
\begin{eqnarray}
\lambda_{{\bf {\tilde q}}E_{2g}}&=&\frac{2\times2\times2 \langle g_{\bf \Gamma}^2\rangle_{F} }
{\hbar\omega_{{\bf {\tilde q}}E_{2g}}N_{\sigma}(\epsilon_f)}
\int_{{\cal F}_{\bf \Gamma}(\epsilon_f)} 
\frac{d^2 {\tilde k}}{\Omega}\times \nonumber \\
& &\times\delta(\epsilon_{{\bf K}+{\bf {\tilde k}} }-\epsilon_f)
\delta(\epsilon_{{\bf K}+{\bf {\tilde k}}+{\bf {\tilde q}} }-\epsilon_f )=\nonumber \\
&=&\frac{8\langle g_{\bf \Gamma}^2\rangle_{F}}{\hbar\omega_{{\bf {\tilde q}}E_{2g}}N_{\sigma}(\epsilon_f)}
\, I_{{\bf {\tilde q}}}\label{eq:pref_I_q}
\end{eqnarray}
where 
\begin{eqnarray}
{\cal F}_{\bf \Gamma}(\epsilon)=
\{{\bf k}|\beta k < \epsilon+\eta\}.
\end{eqnarray}
The $8$ prefactor is the results of having $2$ E$_{2g}$ modes and of having an
identical integral over the second Fermi surface sheet
 at ${\bf K'}$.
The integral $I_{{\bf {\tilde q}}}$ is the so-called nesting factor, defined
as:
\begin{eqnarray}
I_{{\bf {\tilde q}}}=\int_{{\cal F}_{\bf \Gamma}(\epsilon_f)} \frac{d^2 {\tilde k}}{\Omega}
\delta(\epsilon_{{\bf K}+{\bf {\tilde k}} }-\epsilon_f)
\delta(\epsilon_{{\bf K}+{\bf {\tilde k}}+{\bf {\tilde q}} }-\epsilon_f )
\end{eqnarray}

The electron-phonon coupling due to E$_{2g}$ modes is given by
\begin{equation}
\lambda_{{\bf \Gamma}}(\epsilon_f)=N(\epsilon_f)\int_{{\cal F}_{{\bf \Gamma}}(2\epsilon_f)} \frac{d^2 {\tilde q} }{\Omega} 
\lambda_{{\bf {\tilde q}}E_{2g}}=
\frac{2 N(\epsilon_f)\langle g_{\bf \Gamma}^2\rangle_{F}}
{\hbar\omega_{{\bf \Gamma}E_{2g}}}
\label{eq:lambda_G}
\end{equation}
where we have used that 
\begin{eqnarray}
\int_{{\cal F}_{{\bf \Gamma}}(2\epsilon_f)} \frac{d^2 {\tilde q} }{\Omega} I_{\bf {\tilde q}}=
N_{\sigma}^{2}(\epsilon_f)/4
\end{eqnarray}
and we have replaced the E$_{2g}$ phonon frequency with its value at ${\bf \Gamma}$.

Similarly, case (ii) leads to
\begin{eqnarray}
\lambda_{{\bf K}+{\bf {\tilde q}}A_{1}^{'}}
&=&\frac{2\times 2\langle g_{{\bf K}}^2\rangle_{F}}
{\hbar\omega_{{\bf K}+{\bf {\tilde q}} A_{1}^{'}}N_{\sigma}(\epsilon_f)}
\int_{{\cal F}_{{\bf K}}(\epsilon_f)} \frac{d^2 {\tilde k} }{\Omega}
[1-\cos(\theta_{{\bf {\tilde k}},{\bf {\tilde k}}+{\bf {\tilde q}}})]\times\nonumber\\
& &\times\delta(\epsilon_{{\bf K}'+{\bf {\tilde k}} }-\epsilon_f)
\delta(\epsilon_{{\bf K}+{\bf {\tilde k}}+{\bf {\tilde q}}}-\epsilon_f )
=\nonumber \\
&=&\frac{4 \langle g_{{\bf K}}^2\rangle_{F}}{\hbar\omega_{{\bf K}+{\bf {\tilde q}} 
A_{1}^{'}}N_{\sigma}(\epsilon_f)} \, J_{{\bf K}+{\bf {\tilde q}}}
\label{eq:lambda_A1}
\end{eqnarray}
and the additional factor of $2$ is a result of having scattering from
${\cal F}_{\bf K}(\epsilon_f)$ to ${\cal F}_{\bf K'}(\epsilon_f)$ and
vice versa. The quantity, 
\begin{eqnarray}
J_{{\bf K}+{\bf {\tilde q}}}&=&
\int_{{\cal F}_{{\bf K}}(\epsilon_f)} \frac{d^2 {\tilde k} }{\Omega}
[1-\cos(\theta_{{\bf {\tilde k}},{\bf {\tilde k}}+{\bf {\tilde q}}})]\times\nonumber\\
& &\times\delta(\epsilon_{{\bf K}'+{\bf {\tilde k}} }-\epsilon_f)
\delta(\epsilon_{{\bf K}+{\bf {\tilde k}}+{\bf {\tilde q}}}-\epsilon_f )
\label{eq:def_J}
\end{eqnarray}
and its integral over the momentum ${\bf {\tilde q}}$ are evaluated 
in sec. \ref{sec:appendix},
so that the contribution of the A$_{1}^{'}$ mode
to the electron-phonon coupling is:
\begin{eqnarray}
\lambda_{\bf K}(\epsilon_f)=
\int_{{\cal F}_{{\bf K}}(2\epsilon_f)}\frac{d^2 {\tilde q} }{\Omega}  
\lambda_{{\bf K}+\bf {\tilde q}A_{1}^{'}}&=&
\frac{\langle g_{\bf K}^2\rangle_{F} 
N_{\sigma}(\epsilon_f)}{\hbar\omega_{{\bf K} A_{1}^{'}}}
\label{eq:lambda_K}
\end{eqnarray}
where we have approximated 
$\omega_{{\bf K}+{\bf {\tilde q}}  A_{1}^{'}}\approx \omega_{{\bf K} A_{1}^{'}}$.

The total electron phonon coupling is thus:
\begin{equation}
\lambda (\epsilon_f) = N_{\sigma}(\epsilon_f)\left[
\frac{2 \langle g_{\bf \Gamma}^2\rangle_{F}}
{\hbar\omega_{{\bf \Gamma} E_{2g}}} + 
\frac{\langle g_{\bf K}^2\rangle_{F}}{\hbar\omega_{\bf {\tilde q} A_{1}^{'}}}\right]
\label{eq:lambda_graphene}
\end{equation}
which is eq. 1 in ref. \cite{Calandra2005} \\

\section{\label{sec:selfenergy}
Electron self-energy and angle resolved photoemission}

\begin{figure}[t]
\includegraphics[width=0.7\columnwidth]{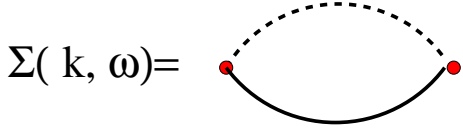}
\caption{Lowest order contribution to the electron self-energy due to the
electron-phonon interaction. The dotted (continuous) line represents the phonon
(electron) self-energy.}
\label{fig:selfediag}
\end{figure}

The lowest contribution to the retarded electron self-energy due to coupling of
$\pi^{*}$ electrons to a phonon mode $\nu$ is illustrated
in Fig. \ref{fig:selfediag}. 
At zero temperature 
direct calculation \cite{Mahan} of the diagram 
gives:
\begin{eqnarray}
\Sigma_{\nu}({\bf k},\epsilon)&=&\sum_{\alpha=\{-1,1\}}
\int_{BZ} \frac{d^2{\bf q}}{\Omega} 
|g_{{\bf k} \pi^{*},{\bf k}+{\bf q} \pi^{*}}^{\nu}|^2 \times\nonumber \\ 
&\times&\left[\frac{\Theta(\alpha\epsilon_f-\alpha\epsilon_{{\bf k}+{\bf q}})}
{\epsilon+i\delta-(\epsilon_{{\bf k}+{\bf q} })
+\alpha\hbar\omega_{{\bf q}\nu}}\right]
\label{eq:selfenergy}
\end{eqnarray}
where $\Theta(x)$ is the Heaviside function.
The imaginary part of eq. \ref{eq:selfenergy} is
\begin{eqnarray}
&&\Sigma_{\nu}^{''}({\bf k},\epsilon)=
-\pi\sum_{\alpha=\{-1,1\}}
\int_{BZ} \frac{d^2{\bf q}}{\Omega} 
|g_{{\bf k} \pi^{*},{\bf k}+{\bf q} \pi^{*}}^{\nu}|^2 \nonumber \\
&\times&\Theta(\alpha\epsilon_f-\alpha\epsilon_{{\bf k}+{\bf q}})
\delta(\epsilon-\epsilon_{{\bf k}+{\bf q} }
+\alpha\hbar\omega_{{\bf q}\nu})=\nonumber \\
&=&-\pi\sum_{\alpha=\{-1,1\}}
\Theta(\alpha\epsilon_f-\alpha\epsilon-\hbar\omega_{\bf q\nu})\nonumber \\
&\times&\int_{BZ} \frac{d^2{\bf k^{'}}}{\Omega} 
|g_{{\bf k} \pi^{*},{\bf k}^{'} \pi^{*}}^{\nu}|^2
 \delta(\epsilon-\epsilon_{{\bf k}^{'}}
+\alpha\hbar\omega_{{\bf k}^{'}-{\bf k}\nu})
\label{eq:Imself_general}
\end{eqnarray}

\begin{figure}[t]
\includegraphics[width=0.9\columnwidth]{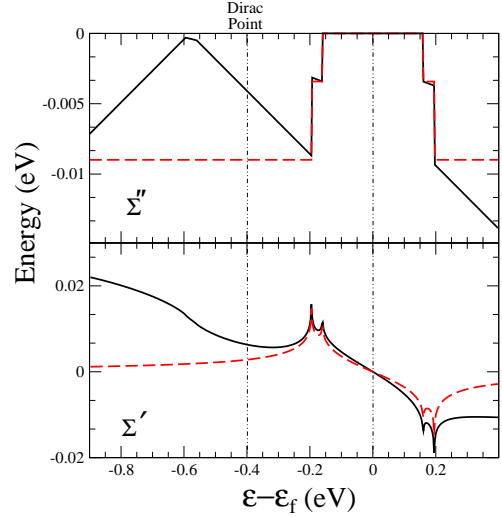}
\caption{(Color online) Real ( $\Sigma^{'}$ ) and 
imaginary $( \Sigma^{''}$ ) parts of the electron self-energy in graphene
(continuous line).
Dashed lines refer to self-energy parts otained using a constant density of
states. The Fermi level is $\epsilon_f=0.4 \,$ eV}
\label{fig:Self_e}
\end{figure}

In angular resolved photoemission (ARPES) experiments the graphene is electron-doped,
so the Fermi level is larger than the Dirac point but it is still in the region where
$\epsilon_{{\bf k}}$ can be considered linear. For a given mode $\nu$ and
a given value of $\alpha$,
the $\delta$-function in eq. \ref{eq:Imself_general} 
restricts the BZ integration to two regions,
close to ${\bf K}$ and to ${\bf K}^{'}$.
The restriction to these regions of k-space and the fact that 
we are interested in the region
of energy-momentum close to the Dirac point, namely
${\bf k}= {\bf K}+{\bf {\tilde k}}$ with $ {\bf {\tilde k}}$ small,
restricts furthermore the integration region. 
Indeed it implies that for small ${\bf {\tilde q}}$:
(i) ${\bf q}={\bf {\tilde q}}$ and (ii) ${\bf q}={\bf K}+{\bf {\tilde q}}$.
Case (i) represents scattering to phonons close to the ${\Gamma}$ point
while (ii) to phonons close to the ${\bf K}$ point.
So the situation is similar to the previous electron-phonon calculation.

The total self-energy, 
$\Sigma^{''}=\sum_{\nu=\{E_{2g},A_{1}^{'}\}} \Sigma_{\nu}^{''}$, 
due to the two E$_{2g}$ phonon modes at ${\bf \Gamma}$ and to the $A_{1}^{'}$
phonon mode at ${\bf K}$ is obtained
substituting Eqs. \ref{eq:g_E2g} and \ref{eq:g_A1} in Eq. \ref{eq:Imself_general},
assuming a constant phonon dispersion around $\Gamma$ and ${\bf K}$
and performing the integration over the BZ, as
\begin{eqnarray}
\Sigma^{''}({\bf {\tilde k}},\epsilon) = 
-\frac{\pi}{2} \sum_{\alpha=\{-1,1\}} \left[\hbar\omega_{{\bf \Gamma}E_{2g}}
\lambda_{\bf \Gamma}(\epsilon-\alpha\hbar\omega_{{\bf \Gamma}E_{2g}})
\right.\times\nonumber \\
\times\Theta(\alpha\epsilon_f-\alpha\epsilon-\hbar\omega_{{\bf \Gamma}E_{2g}})+
\hbar\omega_{\bf K A_{1}^{'}}
\lambda_{\bf K}(\epsilon-\alpha\hbar\omega_{{\bf K}A_{1}^{'}})\times\nonumber \\
\times\left.\Theta(\alpha\epsilon_f-\alpha\epsilon-
\hbar\omega_{{\bf K}A_{1}^{'}})\right]\,\,
\label{eq:graphene_selfe}
\end{eqnarray} 
where $\lambda_{\bf \Gamma}(\epsilon-\hbar\omega_{{\bf \Gamma}E_{2g}})$ and
$\lambda_{\bf K}(\epsilon-\hbar\omega_{{\bf K}A_{1}^{'}})$ are
defined in Eq. \ref{eq:lambda_G} and in Eq. \ref{eq:lambda_K}, respectively.
From Eq. \ref{eq:graphene_selfe} we note that for small ${\bf {\tilde k}}$
the imaginary part of the phonon self-energy is momentum-independent,
so in what follows we drop the ${\bf {\tilde k}}$-label. Using 
numerical values of $\hbar\omega_{{\bf \Gamma}E_{2g}}=0.195 $eV
and $\hbar\omega_{{\bf K}A_{1}^{'}}=0.16 $eV, the 
$\Sigma^{''}(\epsilon)$ is illustrated in Fig. 
\ref{fig:Self_e} (black lines). 

The imaginary part in eq. \ref{eq:graphene_selfe} has to be compared with the 
square well model  which is obtained from Eq. \ref{eq:graphene_selfe}
assuming a constant density of states. This is the commonly used approximation
to interpret ARPES spectra \cite{Grimvall,Cuk_Review}.
The square well model is illustrated
in fig. \ref{fig:Self_e} (red-dashed). In graphene this approximation is 
in principle not allowed due to the behavior of the density of states proportional
to $|\epsilon|$ (see eq. \ref{eq:dos}) . 
The difference between the two models becomes relevant for
energies smaller than or closer to the Dirac point.

The real part of the electron self-energy can be obtained using 
the Kramers-Kronig relations, namely:
\begin{equation}
\Sigma^{'}(\epsilon)=\frac{1}{\pi} {\cal P} \int_{-\infty}^{\infty}
\frac{\Sigma^{''}(\epsilon^{'})}{\epsilon^{'}-\epsilon} \, d\epsilon^{'}
\label{eq:KK}
\end{equation}
If the self-energy
in Eq. \ref{eq:graphene_selfe} is used then $\Sigma^{'}(\epsilon)$
diverges at large $|\epsilon|$ due to the $|\epsilon|$ 
dependence of the density of states. However
this divergence is unphysical since in real graphene the linearity of the $\pi^{*}$
bands and the consequent behavior of the density of states is only up to 
energies of $\approx 1.5$ eV from the Dirac point. Thus for large $|\epsilon|$ the 
imaginary part of the phonon self-energy should be regularized. We adopted the 
following regularization:

\begin{eqnarray}
  \Sigma^{''}_{\rm Reg}(\epsilon)= 
\begin{cases}
\Sigma^{''}(\epsilon) 
                            & \text{if $|\epsilon|<\epsilon_M$} \\
\Sigma^{''}(\epsilon_M) & \text{ if $\epsilon_M<|\epsilon|$ }
\end{cases}
\label{eq:self_regulariz}
\end{eqnarray}
With this assumption, the calculation of the integral in eq. \ref{eq:KK}
leads to:
\begin{eqnarray}
\Sigma^{'}(\epsilon)&=&-N_{\sigma}(\epsilon_f)
\langle g_{\bf\Gamma}^2\rangle\left\{ 
(\epsilon+\hbar\omega_{{\bf \Gamma}E_{2g}})\left[
-2\log\left|\epsilon+\hbar\omega_{{\bf \Gamma}E_{2g}}\right|
+ \right. \right. \nonumber \\
&+&
\left.\left.
\log\left|(\epsilon_M+\epsilon)
(\epsilon_f-\hbar\omega_{{\bf \Gamma}E_{2g}}-\epsilon)\right|
\right]+\right.\nonumber \\
&+&\left.\left.(\epsilon-\hbar\omega_{{\bf \Gamma}E_{2g}})
\log\left|\frac{\epsilon_M-\epsilon}
{\epsilon_f+\hbar\omega_{{\bf \Gamma}E_{2g}}-\epsilon}\right|
\right]\right\}+\nonumber \\
&-&\frac{N_{\sigma}(\epsilon_f)
\langle g_{{\bf K}}^2\rangle}{2}
\left\{ (\epsilon+\hbar\omega_{{{\bf K}}A_{1^{'}}})
\left[
-2\log
\left|\epsilon+\hbar\omega_{{{\bf K}}A_{1^{'}}}\right|\right. \right. \nonumber \\
&+&\left. \left.
\log\left|(\epsilon_M+\epsilon)
(\epsilon_f-\hbar\omega_{{{\bf K}}A_{1^{'}}}-\epsilon)\right|
\right]\right.+\nonumber \\
&+&\left.\left.(\epsilon-\hbar\omega_{{{\bf K}}A_{1^{'}}})
\log\left|\frac{\epsilon_M-\epsilon}
{\epsilon_f+\hbar\omega_{{\bf \Gamma}A_{1^{'}}}-\epsilon}\right|
\right]\right\}+ \nonumber \\
&-&N_{\sigma}(\epsilon_f)\left[
\langle g_{\bf\Gamma}^2\rangle 
(\epsilon-\hbar\omega_{{\bf \Gamma}E_{2g}})+\frac{\langle g_{{\bf K}}^2\rangle}{2}
(\epsilon-\hbar\omega_{{{\bf K}}A_{1^{'}}})\right]\times\nonumber \\
&\times&\left[\log\left|
\frac{\epsilon_M+\epsilon}{\epsilon_M-\epsilon}\right|\right]
\label{eq:real_linear}
\end{eqnarray}
In practical calculations we choose $\epsilon_M=1.5$ eV. 

The real part of the electron self-energy is illustrated in Fig. \ref{fig:Self_e}
and is compared with the real part obtained from the Kramers-Kronig transformation
of $\Sigma^{''}(\omega)$ with a constant density of states.

In ARPES experiment the spectral weight is measured, namely \cite{footnote}
\begin{eqnarray}
A({\bf k},\epsilon)=\frac{-2[\Sigma^{''}(\epsilon)+\eta]}
{\left[\epsilon - \epsilon_{{\bf k}} 
-\Sigma^{'}(\epsilon)\right]^2
+ \left[\Sigma^{''}(\epsilon)+\eta\right]^2}
\label{eq:SW}
\end{eqnarray}
where we allowed for a small constant imaginary
part $\eta$ to eliminate numerical instabilities. Typically
$\eta=10^{-4}$ eV

\section{\label{sec:results}Results}

\begin{figure*}[t]
\includegraphics[height=9.0cm]{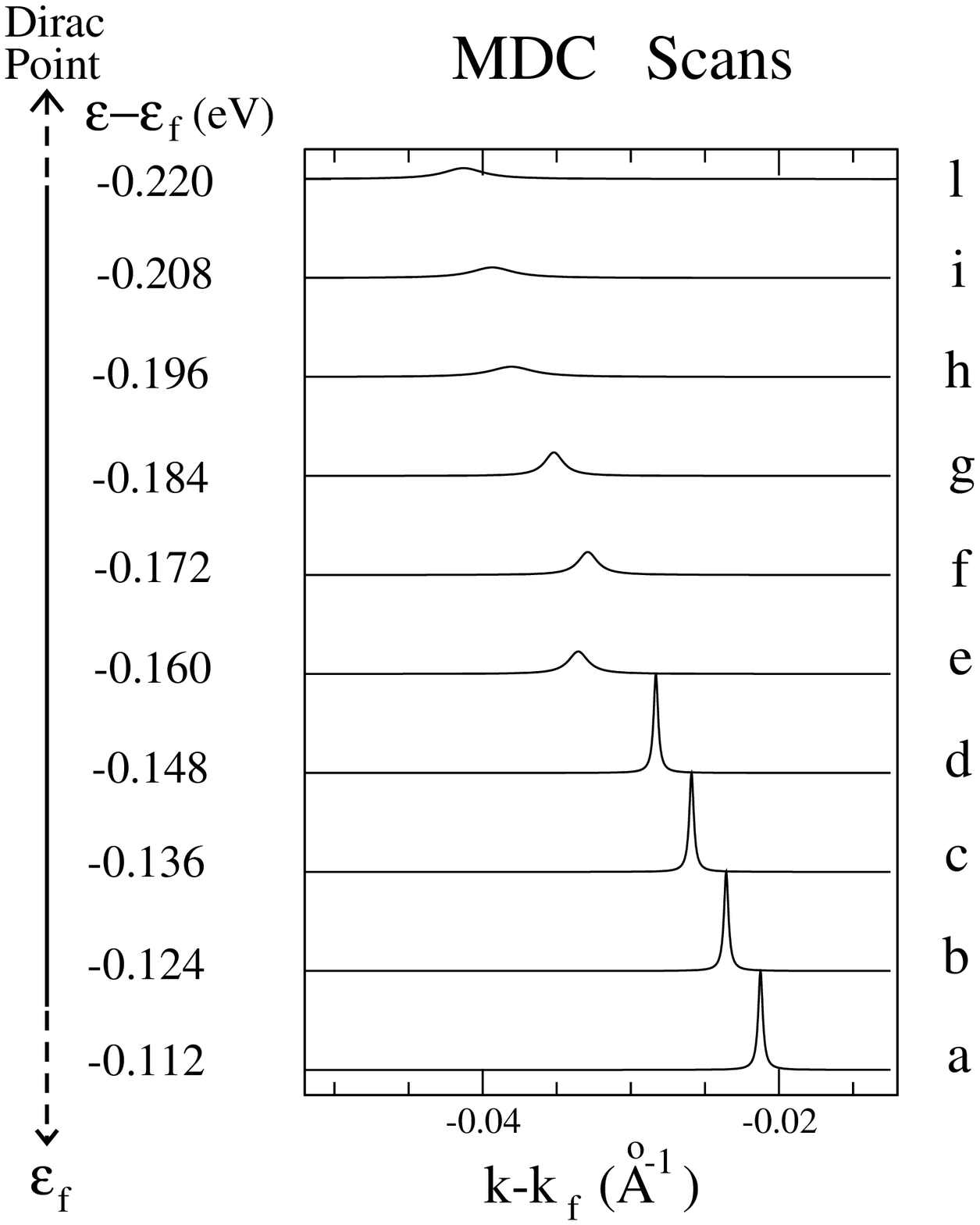}%
\includegraphics[height=9.0cm]{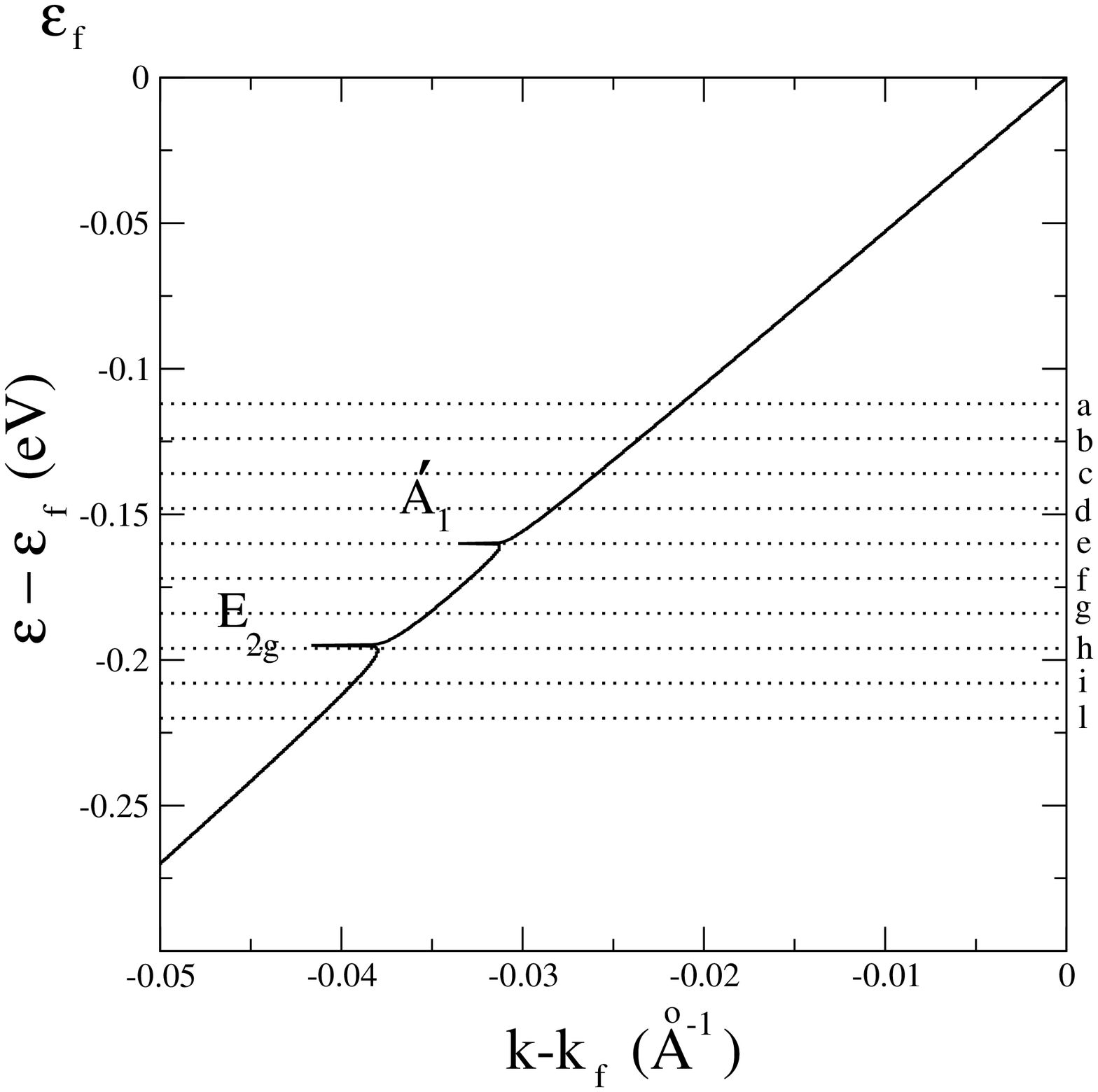}
\caption{MDC scans for different energies (left) and ARPES maximum position
as a function of energy and momentum (right). 
The Fermi level is $\epsilon_f=0.4$ eV, the
experimental resolution is not included in the calculation. A $1\,$ meV broadening  
is used for illustration purposes.}
\label{fig:Arpes-nores}
\end{figure*}

In this section we consider a Fermi level of $\epsilon_f=0.4$ eV measured from
the Dirac point.
Neglecting resolution effects, the spectral function (Eq.  \ref{eq:SW}) 
is shown in figure \ref{fig:Arpes-nores} (left). The scans presented are fixed 
photoelectron energy scans
(the energy value is given on the left of Fig.   \ref{fig:Arpes-nores} (left) )
while the photoelectron momentum is varied (MDC scans). 
In the MDC scans two kinks are present
(compare scans g and h for the first kink and scans e and d for the second),
both in the spectral-weight maximum-position and linewidth.

The behavior of the MDC-maximum-position
as a function of energy and momentum is illustrated in fig. 
\ref{fig:Arpes-nores} (right).
The lower energy kink corresponds to the twofold degenerate $E_{2g}$ mode, while the
higher energy one corresponds to the $A_{1}^{'}$ mode.

\begin{figure}[t]
\includegraphics[width=0.9\columnwidth]{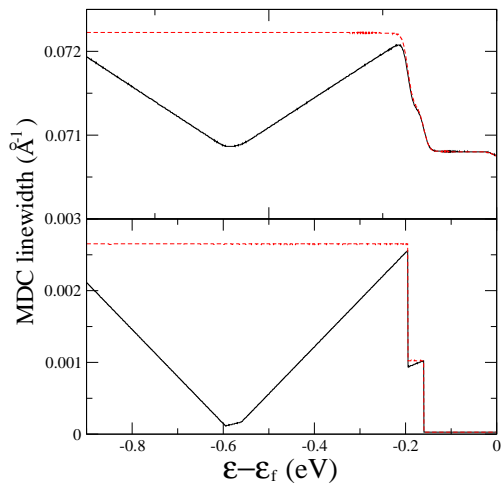}
\caption{(Color online) MDC - linewidth (HWMH) for the two model self-energy.
The curve in upper panel is obtained using a finite resolution, while 
that in the lower one is without resolution effects. In the absence of
resolution effects the linewidth is equal to $-\beta\Sigma^{''}(\epsilon)$
(see fig. \ref{fig:selfediag}).
The Fermi level is $\epsilon_f=0.4 $ eV. }
\label{fig:SW_lw}
\end{figure}

The behavior of the MDC-linewidth, half-width half-maximum (HWMH),
as a function of energy is
shown in fig. \ref{fig:SW_lw}. 
Notice that, in absence of
resolution effects, the linewidth is equal to $-\beta\Sigma^{''}(\epsilon)$.
The behavior of the linewidth is compared with
that of square well model typically used to interpret ARPES data.
The differences are negligible at energies larger than -0.2 eV, but
they are significative at lower energies and in 
particular at the Dirac point. 

To test the robustness of the phonon features against experimental resolution
we introduce the following convoluted spectral weight:
\begin{eqnarray}
A_{\rm exp}({\bf k},\epsilon)&=&f(\epsilon)\int_{-\infty}^{\infty} d\epsilon^{'}
\int d^{3}{\bf k}^{'} \nonumber \\
& &A({\bf k}^{'},\epsilon^{'})G_{\eta_{\epsilon}}(\epsilon^{'}-\epsilon) 
G_{\eta_{\bf k}} ({\bf k}^{'}-{\bf k}) 
\label{eq:resolution}
\end{eqnarray}
where $G_{\eta_x}(x)$ is a Gaussian having full-width $\eta_{x}$ and centered
in $x=0$. The Fermi distribution is indicated with 
\begin{equation}
f(\epsilon)=\frac{1}{\exp{\left[(\epsilon-\epsilon_f)/k_B T\right]}+1}
\end{equation}
This form of the experimental resolution assumes to have decoupled momentum and
energy resolutions.

\begin{figure}[t]
\includegraphics[width=0.9\columnwidth]{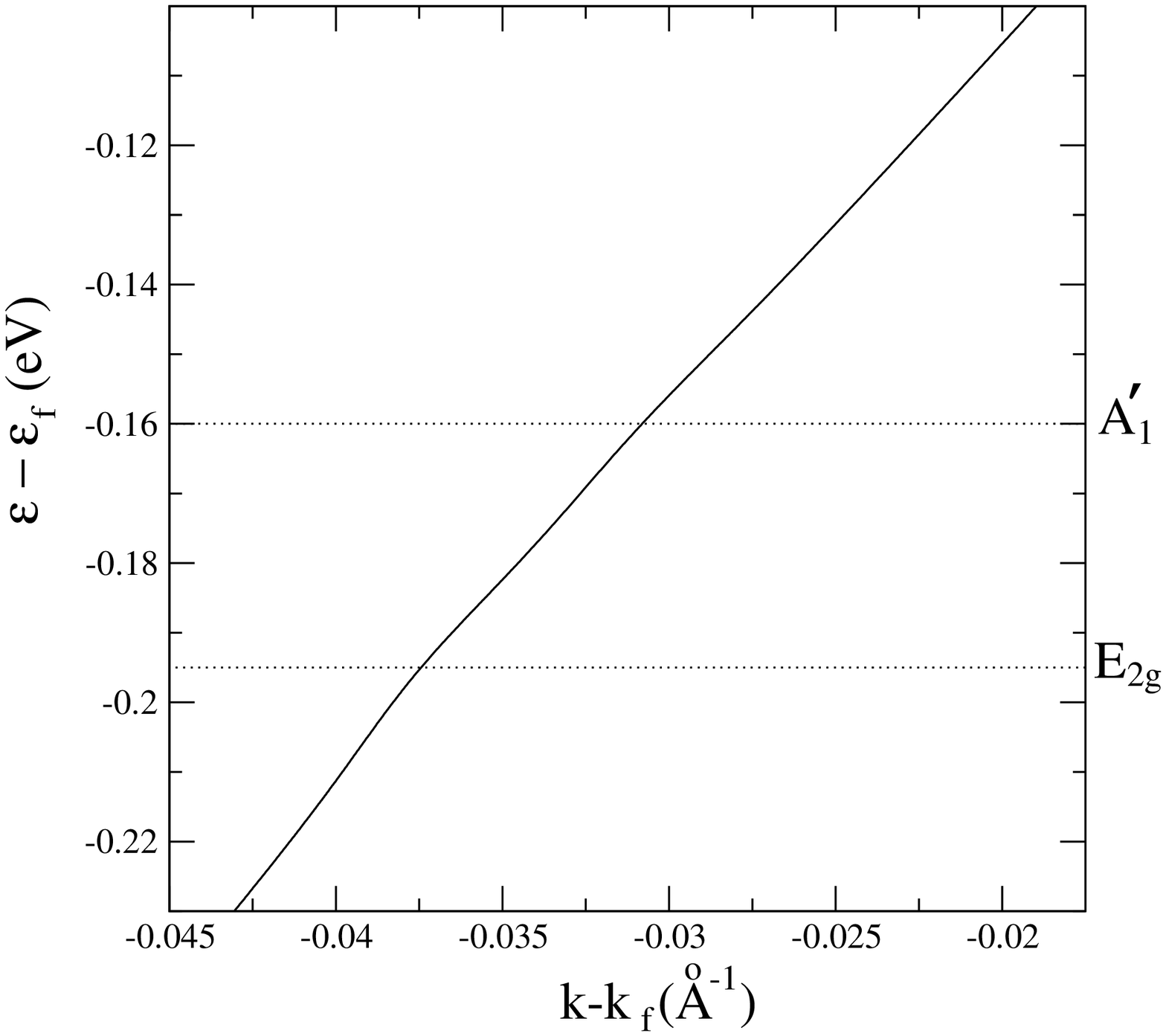}
\caption{Position of the maximum in MDC curves using a finite
resolution. The Fermi level is $\epsilon_f=0.4 $ eV.}
\label{fig:SW_MAX_res}
\end{figure}

We chose $\eta_{\epsilon}=25$ meV and $\eta_{\bf k}=0.1 {\rm \AA}^{-1}$, as
in recent ARPES experiments \cite{Bostwick2007,Bostwick_Review}.
The maximum position in $A_{\rm exp}({\bf k},\epsilon)$ is plotted in Fig.
\ref{fig:SW_MAX_res}.

As it can be seen the experimental resolution substantially
smears out the two kinks so that a very weak kink is visible at -0.2 eV 
while the second one is almost invisible. The kink is substantially smaller
than what detected in experiments \cite{Bostwick2007}. 
The mass enhancement parameter is defined as \cite{Grimvall,Mahan}:
\begin{equation}
\lambda=\left.\frac{\partial 
\Sigma^{'}(\epsilon)}{\partial \epsilon}\right|_{\epsilon=\epsilon_f}.
\end{equation}
Linearizing $\Sigma^{'}(\epsilon)\approx-\lambda\epsilon$,  the spectral-weight
becomes (for $\eta=0$):
\begin{eqnarray}
A({\bf k},\epsilon)\approx\frac{-2\Sigma^{''}(\epsilon) Z^2}
{\left[\epsilon  - Z \epsilon_{{\bf k}} \right]^2
+ \left[\Sigma^{''}(\epsilon)Z\right]^2}
\label{eq:SW_zp}
\end{eqnarray}  
where 
\begin{equation}
Z=\frac{1}{1+\lambda}
\end{equation}
is the quasiparticle weight. From Eq. \ref{eq:SW_zp} one sees that
the quasiparticle state has quasiparticle energy $ Z \epsilon_{{\bf k}}$ and
linewidth $\Sigma^{''}(\epsilon)Z/2$.
In graphene the bands are linear, with $\epsilon_k=\beta k$, so
that the maximum position in the spectral weight at energies higher than
the kink is given by the relation
\begin{equation}
\epsilon_{\bf k}^{\rm max} = \frac{\beta k }{1+\lambda}
\end{equation}
Assuming linear renormalized bands, $\epsilon_{\bf k}=\beta_{\rm ph} k$, 
for energies larger than the kink then 
the following expression for $\lambda$ is obtained:
\begin{equation}
\lambda=\frac{\beta}{\beta_{\rm ph}}-1
\label{eq:lambda_eff}
\end{equation}
Typically $\beta$ is obtained from a linear fit to maximum position
in MDC curves at
energies below the kink (in our case $\beta=5.52 {\rm eV \AA}$ within DFT), 
while $\beta_{\rm ph}$ is obtained  from a linear fit at energies 
higher than the kink but enough below $\epsilon_f$ so that the effects of the 
Fermi function in eq. \ref{eq:resolution} are absent.
  
\begin{figure}[t]
\includegraphics[width=0.9\columnwidth]{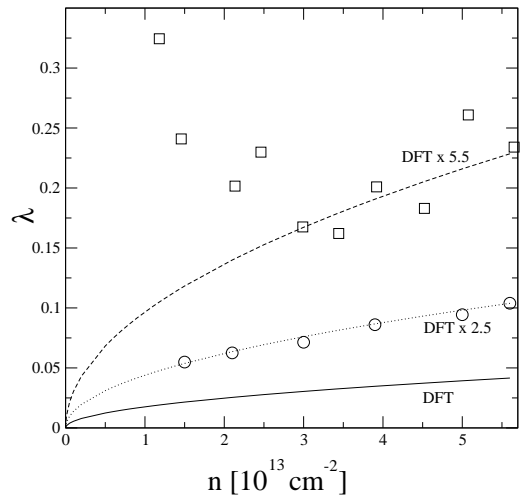}
\caption{Electron-phonon coupling in electron-doped graphene. The continuous line labeled
DFT refers to Eq. \ref{eq:lambda_graphene}, the empty squares are 
experimental data from Fig. 5 in ref. \cite{Bostwick_Review},
while the empty circles represents the 
``apparent'' electron-phonon coupling extracted from fits to the calculated
ARPES spectra using Eq. \ref{eq:lambda_eff} and including resolution effects. 
The curves DFT$\times 5.5$ and 
DFT$\times 2.5$ indicates Eq. \ref{eq:lambda_eff} with 
multiplied by the prefactors $5.5$ and $2.5$ respectively.}
\label{fig:Eli_plot}
\end{figure}

When a finite resolution is used, the result is substantially affected.
Indeed, we find that a linear fit in the energy window
from  $-0.195 \,{\rm eV} < \epsilon -\epsilon_f < -0.04 \,{\rm eV}$ leads
to values of the electron-phonon coupling which are 
a factor 2.5 than Eq. \ref{eq:lambda_graphene}, as shown in Fig. 
\ref{fig:Eli_plot}.
A similar fitting procedure is used in experiments (note that in
experiments only the kink
due to the E$_{2g}$ phonon is visible). A comparison with 
available experimental data (see fig. 5 in Ref. \cite{Bostwick_Review})
is shown in fig. \ref{fig:Eli_plot}. A clear disagreement between 
theory and experiments is still present, even if it is reduced by resolution
effects.

For this reason, we believe, the determination of $\lambda$ using
Eq. \ref{eq:lambda_eff} is affected by a large error. The error is
due to the difficulties in the determination of $\beta_{\rm ph}$  
generated by the non-linearity in $\Sigma^{'}(\epsilon)$ near the kink.
The non-linearity, when convoluted with a finite resolutions, results
in a quasi-linear behavior with an ``apparent'' enhanced electron-phonon
coupling.

\section{\label{sec:experimentals}Electron relaxation times in graphite and graphene}

\subsection{Electron-doped graphene}

It is interesting to compare the ARPES-measured self-energy imaginary part with what 
detected by alternative experimental techniques.
In electron-doped graphene the electron relaxation time has been determined
experimentally by conductivity/mobility data \cite{Zhang2005} and 
by angular resolved 
photoemission measurements \cite{Bostwick2007}.
The mobility measurement detect the scattering time of electrons at 
energies $|\epsilon - \epsilon_f|<k_B T$, where $k_B$ is the Boltzmann
constant and T is the temperature
at which the experiment is performed ($T\approx 300$ K ). As shown in 
appendix \ref{sec:app-mobility}, this leads to electron scattering time
of the order of 
\begin{equation}
\tau = 0.35  {\rm ps}  \,\,\,\,\,\,\, {\rm (from\,\, mobility)} 
\label{eq:graphene-taumu}
\end{equation}
for $\epsilon_f>0.2 $eV. Since the Debye temperature of the optical phonon in
graphene is much larger then 300 K, the scattering in the mobility measurements
is mainly due to defects and acoustic phonons.

ARPES measures the electron self-energy at 
photo-emitted electron-energies $\epsilon$.
The imaginary part of the electron self-energy 
is related to the electron scattering-time by the
relation $\tau(\epsilon) = \hbar/[2\Sigma^{''}(\epsilon)]$.
From the data in Refs. \cite{Bostwick2007,Zhou_Private}
for $|\epsilon-\epsilon_f| \approx k_B T$ we obtain
\begin{equation}
\tau \approx 3.5 {\rm fs} \,\,\,\,\,\,\, {\rm (from\,\, ARPES)} 
\label{eq:graphene-tauarpes}
\end{equation}
which is two order of magnitudes smaller than Eq. \ref{eq:graphene-taumu}.

\subsection{Graphite}

In graphite the electron scattering time has been measured by two different 
experimental techniques; 
(i) Femptoseconds time-resolved spectroscopy \cite{Moos} and (ii) ARPES.

From femptoseconds time-resolved spectroscopy \cite{Moos}:
\begin{eqnarray}
\tau &\approx& 0.2 {\rm ps \,\,\,\,\,for\,\,\,} |\epsilon-\epsilon_f|= 0.25\, {\rm eV} \,\,\,\,\,\,\,  \nonumber \\
\tau &\approx& 0.1 {\rm ps \,\,\,\,\, for\,\,\,} |\epsilon-\epsilon_f|= 0.50\, {\rm eV} \,\,\,\,\,\,\, \nonumber \\
& &{\rm ( from\,\, femtoseconds  } \nonumber \\
& & {\rm time-resolved\,\, spectroscopy)} 
\end{eqnarray}

Similar to what happens in graphene,
ARPES measurements \cite{Sugawara,ZhouPRB2005,Zhou2006NatPhys,Zhou_Annals} lead
to a relaxation time which are two order of magnitudes times smaller than
what obtained from the femptoseconds photoemission spectroscopy. For example,
from the measured ARPES linewidth in Ref. \cite{Zhou_Annals} (see Fig. 10 c),
we obtain:
\begin{eqnarray}
\tau &\approx& 4.7  {\rm fs  \,\,\,\,\,for\,\,\,} |\epsilon-\epsilon_f|= 0.25\, {\rm eV} \,\,\,\,\,\,\,  \nonumber \\
\tau &\approx& 3.4 {\rm fs \,\,\,\,\, for\,\,\,} |\epsilon-\epsilon_f|= 0.50\, {\rm eV} \,\,\,\,\,\,\, \nonumber \\
& &  \,\,\,\,\, {\rm (from\,\, ARPES)} 
\end{eqnarray}

\section{Conclusions\label{sec:conclusions}}

In this work we calculated the electron-phonon coupling parameter and the
electron-phonon coupling contribution to the electron self-energy in 
doped graphene. 
From the electron self-energy we obtained the spectral-weight function and
the ARPES spectra. 

The ARPES spectra
as a function of momentum and energy displays two kinks. The kinks are
at energies $\epsilon-\epsilon_f \approx -0.195$ eV and
$\epsilon-\epsilon_f \approx -0.16$ eV, where $\epsilon_f$ is the Fermi level.
The two kinks are due to coupling to the twofold degenerate E$_{2g}$ mode and
to the A$_{1}^{'}$ mode respectively. The MDC-linewidth 
as a function of energy is discontinuous (jump) 
at E$_{2g}$ and A$_{1}^{'}$ phonon energies.

Comparing the calculated electron-phonon coupling with that extracted 
from ARPES experiments we found that, for large enough electron-doping,
the latter is roughly a factor 5.5 larger than the former, 
as suggested in ref. \cite{Bostwick_Review}. 
We partially solved this contradiction by including finite resolution
effects. Indeed, in experiments the 
electron-phonon coupling is determined from the ratio of the electron-velocities
at higher and lower energies respect to the kink. 
The velocities are obtained from the slopes 
of the maximum position
of the ARPES spectra as a function of energy and momentum.
We find that the slope above the kink is substantially affected by the 
presence of a finite resolution and the extracted values of the electron-phonon
coupling are $\approx 2.5$ larger than what obtained without any
resolution effect. 
Thus when comparing calculated spectra with the inclusion of finite resolution
effects to ARPES experiments \cite{Bostwick2007,Bostwick_Review}
we remark that the measured electron-phonon coupling is
still a factor of 2.2 larger than the calculated one \cite{Calandra2005}.
Thus this work shows once more \cite{Valla} the importance of including
resolution effects to correctly describe ARPES data.

Finally, from the imaginary part of the
electron self-energy we obtain the electron-relaxation time.
The calculated electron relaxation time is in good agreement with 
mobility data on electron-doped graphene and is of the same order
of magnitude of the electron relaxation time obtained from 
conductivity and femtoseconds time-resolved spectroscopy
measurements in graphite. However this is in strong disagreement
with ARPES measurements, being the ARPES relaxation times, both
in graphene and graphite, almost
two order of magnitudes smaller. This discrepancy essentially
reflects the disagreement in the measured and calculated electron-phonon 
coupling. 
The aforementioned disagreement in the electron self-energies
is even more surprising when considering 
that previous DFT calculations of the phonon
self-energy in graphene, graphite and nanotubes 
were found to be in perfect agreement with experimental data
for what concerns phonon dispersion and phonon lifetimes,
\cite{Pisana,Lazzeri2006, LazzeriPRB2006}.
Since the electron and phonon self-energies involve the same vertex,
and thus the same matrix elements, a good agreement would be
expected even for the electron self-energies too.

\section{Acknowledgments}

We acknowledge illuminating discussions with Olle Gunnarsson, Eli Rotenberg,
Aaron Bostwick, Jessica McChesney and Shuyun Zhou.
Calculations were performed at the IDRIS supercomputing center (project 071202).

\section{Appendix}

\subsection{Evaluation of the integral $J_{{\bf K}+{\bf {\tilde q}}}$}\label{sec:appendix}
In Eq. \ref{eq:lambda_A1},
 $\theta_{{\bf {\tilde k}},{\bf {\tilde k}}+{\bf {\tilde q}}}=
2\theta_{{\bf {\tilde k}}{\bf {\tilde q}}}-\pi$, so
that 
$1-\cos(\theta_{{\bf {\tilde k}},{\bf {\tilde k}}+{\bf {\tilde q}}})=
2\cos^2(\theta_{{\bf {\tilde k}}{\bf {\tilde q}}})$ and the
integral $J_{{\bf K}+{\bf {\tilde q}}}$ (see Eq. \ref{eq:def_J} )
can be evaluated as:
\begin{eqnarray}
J_{{\bf K}+{\bf {\tilde q}}}&=&
\int_{0}^{2\pi}d\theta_{{\bf {\tilde k}}{\bf {\tilde q}}}
\int \frac{d{\tilde k}}{\Omega}\,
2{\tilde k}\cos^{2}(\theta_{{\bf {\tilde k}}{\bf {\tilde q}}})
\delta(\beta {\tilde k}-\beta k_f)\times \nonumber \\
& &\times
\delta(\beta|{\bf {\tilde k}}+{\bf {\tilde q}}|-\beta k_f )=\nonumber \\
&=&\frac{2\epsilon_f}{\Omega \beta^3}\int_{0}^{2\pi}d\theta_{{\bf {\tilde k}}{\bf {\tilde q}}}
\,\cos^{2}(\theta_{{\bf {\tilde k}}{\bf {\tilde q}}})\times \nonumber \\
& &\times\delta(\sqrt{ k_{f}^2 +  {\tilde q}^2 + 
  2{\tilde k}{\tilde q}
\cos(\theta_{{\bf {\tilde k}}{\bf {\tilde q}}})}-k_f)=\nonumber \\
&=&\frac{2k_f}{\Omega \beta^2}\sum_{\alpha=1,2}
\int_{-1}^{1}d(\cos(\theta_{{\bf {\tilde k}}{\bf {\tilde q}}}^{\alpha}))
\delta(\cos(\theta_{{\bf {\tilde k}}{\bf {\tilde q}}}^{\alpha})+
\frac{{\tilde q}}{2k_f})\times
\nonumber \\
& &\times\frac{\cos^{2}(\theta_{{\bf {\tilde k}}{\bf {\tilde q}}}^{\alpha})
\sqrt{k_f^2 + {\tilde q}^2 + 
2k_f{\tilde q}
\cos(\theta_{{\bf {\tilde k}}{\bf {\tilde q}}}^{\alpha})}}
{k_{f} {\tilde q}|\sin(\theta_{{\bf {\tilde k}}{\bf {\tilde q}}}^{\alpha})|}
=\nonumber \\
&=&\frac{2}{\Omega \beta^2}
\frac{{\tilde q}/2k_f}{ \sqrt{1-\frac{{\tilde q}^2}{4k_f^2}}}
\end{eqnarray}
Furthermore, the integral of such a quantity is:
\begin{eqnarray}
A&&=\int_{{\cal F}_{\bf K}(2\epsilon_f)}\frac{d^2 {\bf {\tilde q}}}{\Omega}\,
J_{{\bf K}+{\bf {\tilde q}}}=\nonumber \\
&&=\frac{2}{\Omega^2 \beta^2}\int_{{\cal F}_{\bf K}(2\epsilon_f)} d^2 {\bf {\tilde q}}
\frac{q/2k_f}{\sqrt{1-\frac{q^2}{4k_f^2}}}=\nonumber \\
&&=\frac{8 k_f^2}{\Omega^2 \beta^2}\int_{0}^{2\pi} d\theta \int_{0}^{1} dy
\frac{y^2}{\sqrt{1 - y^2}}=\nonumber \\
&&=\frac{2\pi 8 k_f^2}{\Omega^2 \beta^2}
\left[-\frac{y\sqrt{1-y^2}}{2}+\frac{1}{2}{\rm arcsin}(y)\right]_{0}^{1}=\nonumber \\
&&=
\frac{16\pi  k_f^2}{\Omega^2 \beta^2}\frac{\pi}{4}=
\frac{16\pi^2  \epsilon_f^2}{\Omega^2 \beta^4}\frac{1}{4}=
\frac{N_{\sigma}^{2}(\epsilon_f)}{4}
\label{eq:int_JK}
\end{eqnarray}

\subsection{From mobility to electron relaxation time \label{sec:app-mobility}} 

The conductivity tensor is \cite{Ashcroft}
\begin{eqnarray}
\sigma=2e^2\sum_{\nu}\int \frac{d^2k}{(2\pi)^2} 
\tau_{\nu}({\bf k}) {\bf v}_{\nu}({\bf k})\cdot {\bf v}_{\nu}({\bf k})
\left(-\frac{\partial f}{\partial \epsilon}\right)_{\epsilon=\epsilon_{\nu}({\bf k})}
\label{eq:sigma-general}
\end{eqnarray}
Eq. \ref{eq:sigma-general} should then divided by two since we are interested 
in only one component of the conductivity tensor. Moreover we only consider
intraband transition since they are the only relevant at low temperature.
The graphite bands are linear so that $\epsilon_k = \pm \beta k = \pm \hbar v_f k$ where
$k$ measure the distance from the ${\bf K}-$point. At zero temperature only electrons 
on the Fermi surface contribute to the integral thus, assuming a constant 
relaxation time, one gets:
\begin{eqnarray}
\sigma&\approx& e^2\sum_{\nu} v_f^2 \tau \,\,
2\int_{{\cal F}_{\bf K}} \frac{dk k}{(2\pi)^2}
\int_{0}^{2\pi} d\theta \delta (\epsilon_f- \epsilon_{{\bf k}\nu})=\nonumber \\
&=&  e^2\sum_{\nu} \frac{v_f^2 \tau}{\hbar^2 v_f^2 }\frac{ \epsilon_f }{\pi}
=\nonumber \\
&=& \frac{ e^2 \tau \epsilon_f}{\hbar^2  \pi}
\end{eqnarray}
so that
\begin{eqnarray}
\tau=\frac{\hbar^2  \pi \sigma}{ e^2  \epsilon_f}
\end{eqnarray}
The conductivity can be written as a function of the mobility as
$\sigma=\delta e \mu$, where $\delta$ is the number of electrons
participating in conduction per surface area.
Using Eq. 1 in Ref. \cite{Lazzeri2006}
$\delta=\epsilon_f^2/(\pi \beta^2)$ and 
$\sigma=\epsilon_f^2 e \mu/(\pi \beta^2)$. Thus the relaxation time
becomes:
\begin{eqnarray}
\tau=\frac{\hbar^2  \pi \delta \mu }{ e  \epsilon_f}
=\frac{\hbar^2  \epsilon_f \mu}{e \beta^2}
\end{eqnarray}
where $\epsilon_f$ is expressed in eV and $\mu$ in 
${\rm cm}^2/({\rm Volt}\times{\rm sec})$.  Expressing
the mobility in ${\rm m}^2/({\rm Volt}\times{\rm sec})$
one gets
\begin{equation}
\tau=1.47\times 10^{-12} \epsilon_f \mu=1.47\times \epsilon_f \mu \, {\rm ps}
\end{equation}
Using the values of Ref. \cite{Zhang2005} for mobility one gets at large doping 
($\epsilon_f > .18$ eV) values of the order of $0.35 $ps.

%Unused bibitems

%\bibitem{Philipp} H. R. Philipp, Phys. Rev. B {\bf 16}, 2896 (1977)
%Unused bibitems

%\bibitem{Reynolds} J. M. Reynolds, H. W. Hemstreet, and T. E. Leinhardt
%Phys. Rev. {\bf 91}, 1152 (1953)
%
\end{document}